\documentclass[final,5p,times,twocolumn]{elsarticle}

\usepackage{graphicx}
\graphicspath{{./graphics/}}

\usepackage{amsmath}
\usepackage{amssymb}
\usepackage{url}
\usepackage{float}
\usepackage{tikz}

\ifcsname italic \endcsname%
\else%
  
\fi

\newcommand{\pd}[2]{\frac{\partial #2}{\partial #1}}
  
\journal{Nuclear Instruments and Methods A}

\begin{document}

\begin{frontmatter}



\title{Measurement and tricubic interpolation of the magnetic field \\for the OLYMPUS experiment}


\author[mit]{J.~C.~Bernauer}

\author[hampton]{J.~Diefenbach\fnref{fn1}}
\fntext[fn1]{Currently with Johannes Gutenberg-Universit\"at, Mainz, Germany}

\author[yerphi]{G.~Elbakian}

\author[pnpi]{G.~Gavrilov}

\author[desy]{N.~Goerrissen}

\author[mit]{D.\ K.~Hasell}

\author[mit]{B.\ S.~Henderson\corref{cor}}
\ead{bhender1@mit.edu}

\author[desy]{Y.~Holler}

\author[yerphi]{G.~Karyan}

\author[desy]{J. Ludwig}

\author[yerphi]{H.~Marukyan}

\author[pnpi]{Y.~Naryshkin}

\author[mit]{C.~O'Connor}

\author[mit]{R.\ L.~Russell}

\author[mit]{A.~Schmidt}

\author[desy]{U.~Schneekloth}

\author[pnpi]{K.~Suvorov}

\author[pnpi]{D.~Veretennikov}

\cortext[cor]{Corresponding author}

\address[mit]{Massachusetts Institute of Technology, Laboratory for Nuclear Science, Cambridge, MA, USA}
\address[hampton]{Hampton University, Hampton, VA, USA}
\address[yerphi]{Alikhanyan National Science Laboratory (Yerevan Physics
Institute), Yerevan, Armenia}
\address[pnpi]{Petersburg Nuclear Physics Institute, Gatchina, Russia}
\address[desy]{Deutsches Elektronen-Synchrotron DESY, Hamburg, Germany}

\begin{abstract}
The OLYMPUS experiment used a 0.3~T toroidal magnetic spectrometer to measure the momenta of outgoing 
charged particles. In order to accurately determine particle trajectories, knowledge of the magnetic 
field was needed throughout the spectrometer volume. For that purpose, the magnetic field was 
measured at over 36,000 positions using a three-dimensional Hall probe actuated by a system of translation tables. 
We used these field data to fit a numerical magnetic field model, which could be employed to calculate 
the magnetic field at any point in the spectrometer volume. Calculations with this model were computationally
intensive; for analysis applications where speed was crucial, we pre-computed the magnetic
field and its derivatives on an evenly spaced grid so that the field could be interpolated between grid points. 
We developed a spline-based interpolation scheme suitable for SIMD implementations, with a memory layout chosen 
to minimize space and optimize the cache behavior to quickly calculate field values. This scheme requires only 
one-eighth of the memory  needed to store necessary coefficients compared with a previous scheme \cite{NME:NME1296}. 
This method was accurate for the vast majority of the spectrometer volume, though special fits and representations 
were needed to improve the accuracy close to the magnet coils and along the toroid axis.

\end{abstract}


\end{frontmatter}



\section{Introduction}

\begin{figure}[hptb]
  
  \begin{tikzpicture}
    \node[anchor=south west,inner sep=0] (image) at (0,0)
         {
           \includegraphics[width=\columnwidth,clip=true,trim=0 5.5cm 0 4cm]{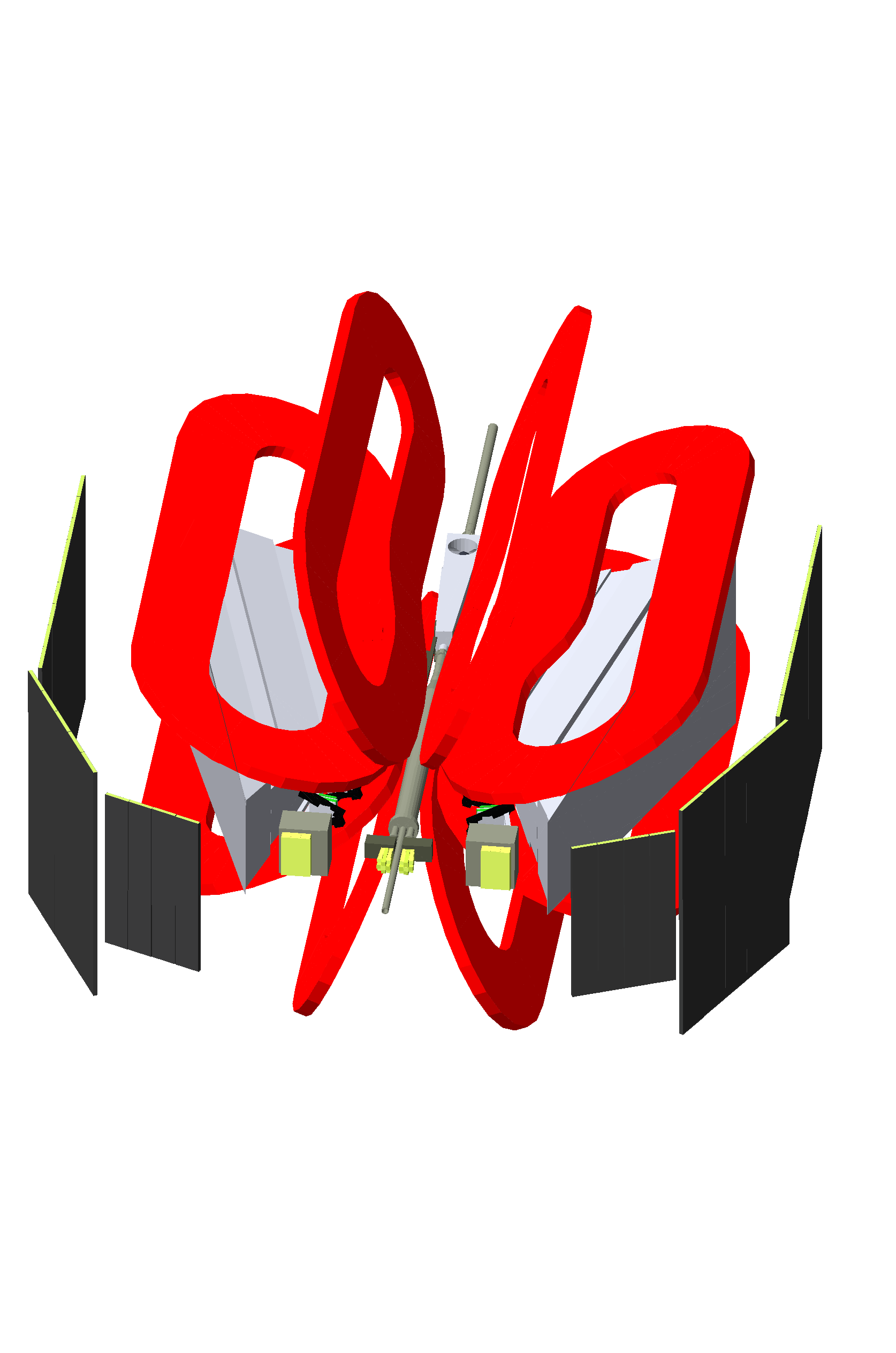}
         };
         \begin{scope}[x={(image.south east)},y={(image.north west)},scale=1.15]
           \node[align=center,black] at (.11,.68) {$z$};
           \node[align=center,black] at (.12,.82) {$y$};
           \node[align=center,black] at (.18,.75) {$x$};
           \node[align=left,black] at (.72,.69) {target chamber};
           \node[align=center,black] at (.518,.805) {beam direction};
           \draw[arrows=->,draw=black,thick] (.1,.73) -- (.1,.83) node[midway,above]{};
           \draw[arrows=->,draw=black,thick] (.1,.73) -- (.2,.73) node[midway,above]{};
           \draw[arrows=->,draw=black,thick] (.1,.73) -- (0.085,.67) node[midway,above]{};
           \draw[arrows=-,draw=black]  (.1,0.01) -- (.1,0.04) node[midway,above]{};
           \draw[arrows=-,draw=black]  (.711,0.01) -- (.711,0.04) node[midway,above]{};
           \node[align=center,black] at (.406,0.02) {5~m};
           \draw[arrows=<-,draw=black,thick] (.1,.02) -- (.37,.02) node[midway,above]{};
           \draw[arrows=<-,draw=black,thick] (.711,.02) -- (.442,.02) node[midway,above]{};
           \draw[arrows=->,draw=black,thick] (.506,.785) -- (.482,.660) node[midway,above]{};
           \draw[arrows=->,draw=black,thick] (.606,.69) -- (.445,.495) node[midway,above]{};
         \end{scope}
  \end{tikzpicture}
  \caption{This schematic shows how the eight magnet coils are situated around the
    OLYMPUS detectors.
    \label{fig:olympus}}
\end{figure}
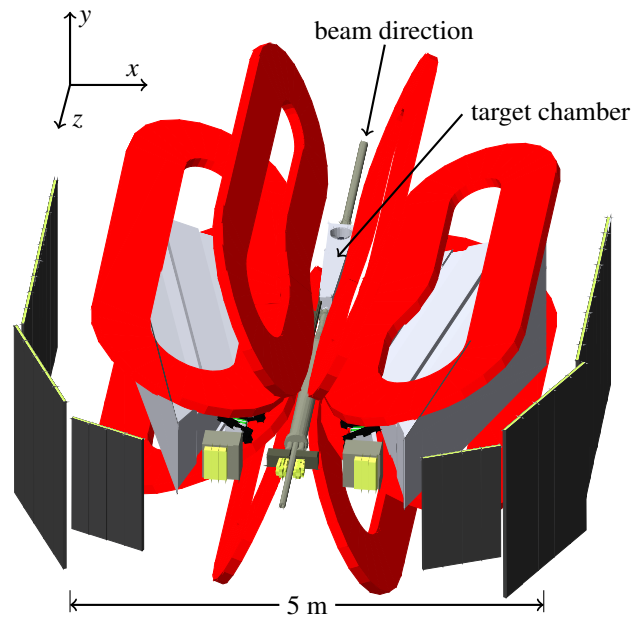

OLYMPUS is a particle physics experiment comparing the elastic cross section
for positron-proton scattering to that of electron-proton scattering \cite{Milner:2013daa}. This 
measurement has been of interest recently because it tests the hypothesis that 
hard two-photon exchange is responsible for the proton form factor discrepancy \cite{Guichon:2003qm,Blunden:2003sp}.
OLYMPUS took data in 2012 and 2013 at the DORIS storage ring at DESY, in Hamburg, 
Germany. During data taking, beams of electrons and positrons were directed 
through a windowless hydrogen gas target. The scattered lepton and recoiling 
proton from elastic scattering events were detected in coincidence with a toroidal 
magnetic spectrometer. The spectrometer's support structure, magnet, and several 
subdetectors were originally part of the BLAST experiment \cite{Hasell:2009zza}. 
Several new detectors were specially built for OLYMPUS to serve as luminosity monitors. 
A schematic of the apparatus is shown in Figure \ref{fig:olympus}.

The OLYMPUS spectrometer used a magnetic field for two purposes.
First, the field produced curvature in the trajectories of charged particles
so that the detectors could measure their momentum. Typical momenta of particles
from elastic scattering reactions ranged from $0.2$ to $2$~GeV$/c$,
corresponding to sagittas as small as $5$~mm in the OLYMPUS tracking detectors. Secondly, the magnet acted
as a filter, preventing low-energy charged particles (from background processes
like M\o ller or Bhabha scattering) from reaching the tracking detectors. 

\begin{figure}[hptb]
\includegraphics[width=\columnwidth]{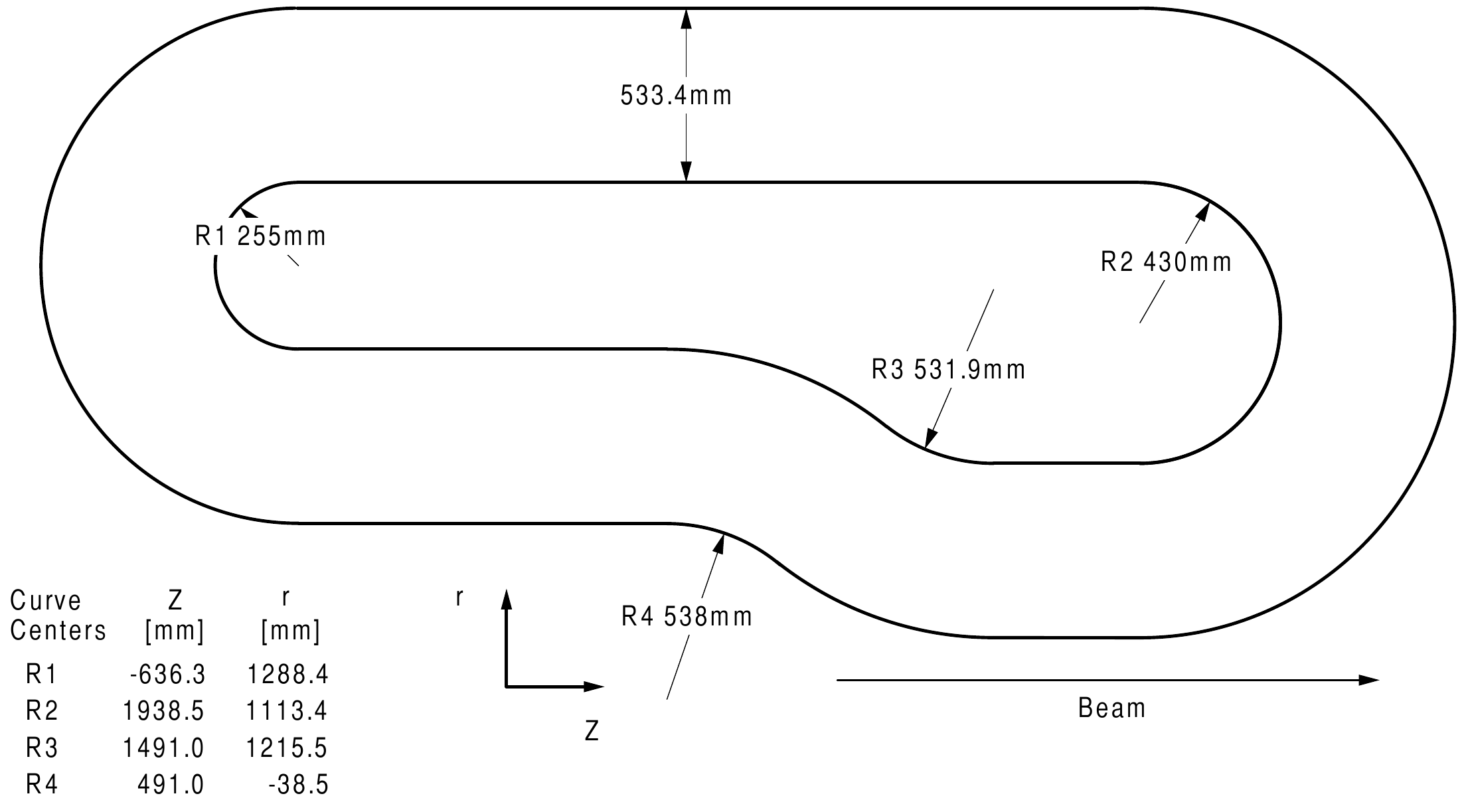}
\caption{Individual magnet coils were narrower at the upstream end to
accommodate the target chamber.
\label{fig:coil}}
\end{figure}

The OLYMPUS magnet consisted of eight coils, identical in shape, each with
26 windings of hollow copper bars potted together with epoxy. 
The coil shape is shown in Figure \ref{fig:coil}. 
The coils were arranged to produce a toroidal field, 
with the beamline passing down the symmetry axis of the toroid. During 
OLYMPUS running, the coils carried 5000~A of current, which produced a
peak field of approximately 0.3~T. 

Knowledge of the spectrometer's magnetic field was necessary for reconstructing 
particle trajectories through the OLYMPUS spectrometer. However, calculating
the field using the design drawings and the Biot-Savart law was not feasible for two reasons. First,
the positions of the copper bars within the epoxy were not well known. Secondly,
the coils were observed to deform slightly when current passed through them due to magnetic forces. 
Instead, at the conclusion of data taking, extensive field measurements
were made of the magnet in situ. A measurement apparatus, consisting of a 
three-dimensional Hall probe actuated by a system of translation tables, 
was used to measure the magnetic field vector at over 36,000 positions in
both the left- and right-sector tracking volumes.

This paper presents both the measurement technique and the subsequent
data analysis used to characterize the field of the OLYMPUS magnet.
Section \ref{sec:measurements} describes the measurement apparatus,
the measurement procedure, and the techniques used to establish the Hall probe 
position. Section \ref{sec:coilfitting} describes how we fit the magnetic field
data with a numerical field model to allow us to calculate the field at positions
we did not measure. Section \ref{sec:interpolation} describes the special interpolation
scheme we developed to facilitate rapid queries of the magnetic field. Finally, Section
\ref{sec:special} describes the special modifications to our field model that were
needed in two regions where the model did not perform adequately.

\section{Coordinate System}

This paper makes frequent references to positions and directions in
the standard OLYMPUS coordinate system. In this system, the $x$-axis points 
left from the beam direction, the $y$-axis points up, 
and the $z$-axis points in the direction of the beam. The 
coordinate origin is taken to be the center of the target. OLYMPUS has 
two sectors of detectors, which lie to the left ($x>0$) and right ($x<0$) 
of the beamline, centered on the $y=0$ plane. Since the magnet has toroidal 
symmetry, it is sometimes convenient to work with cylindrical coordinates.
We use $r$ to refer to the radius from the $z$ axis and 
$\phi$ to refer to the azimuthal angle from the $xz$ plane. For example, 
a point on the positive $y$ axis lies at $\phi=90^\circ$.

\section{Field Measurements}
\label{sec:measurements}

The magnetic field measurements at OLYMPUS were more involved than those 
made during the BLAST experiment, detailed in a previous article
\cite{Dow:2009zz}. Like at BLAST, preliminary field measurements along
the beamline were made to align the coils during the toroid's assembly; 
in addition, a detailed survey was made after data taking was complete. 
This was important because OLYMPUS compared scattering with electrons to scattering
with positrons; the magnetic field introduces differences in trajectories
between the two species. Field inaccuracies directly contribute to the systematic error.

\begin{figure}[hptb]
\includegraphics[width=\columnwidth]{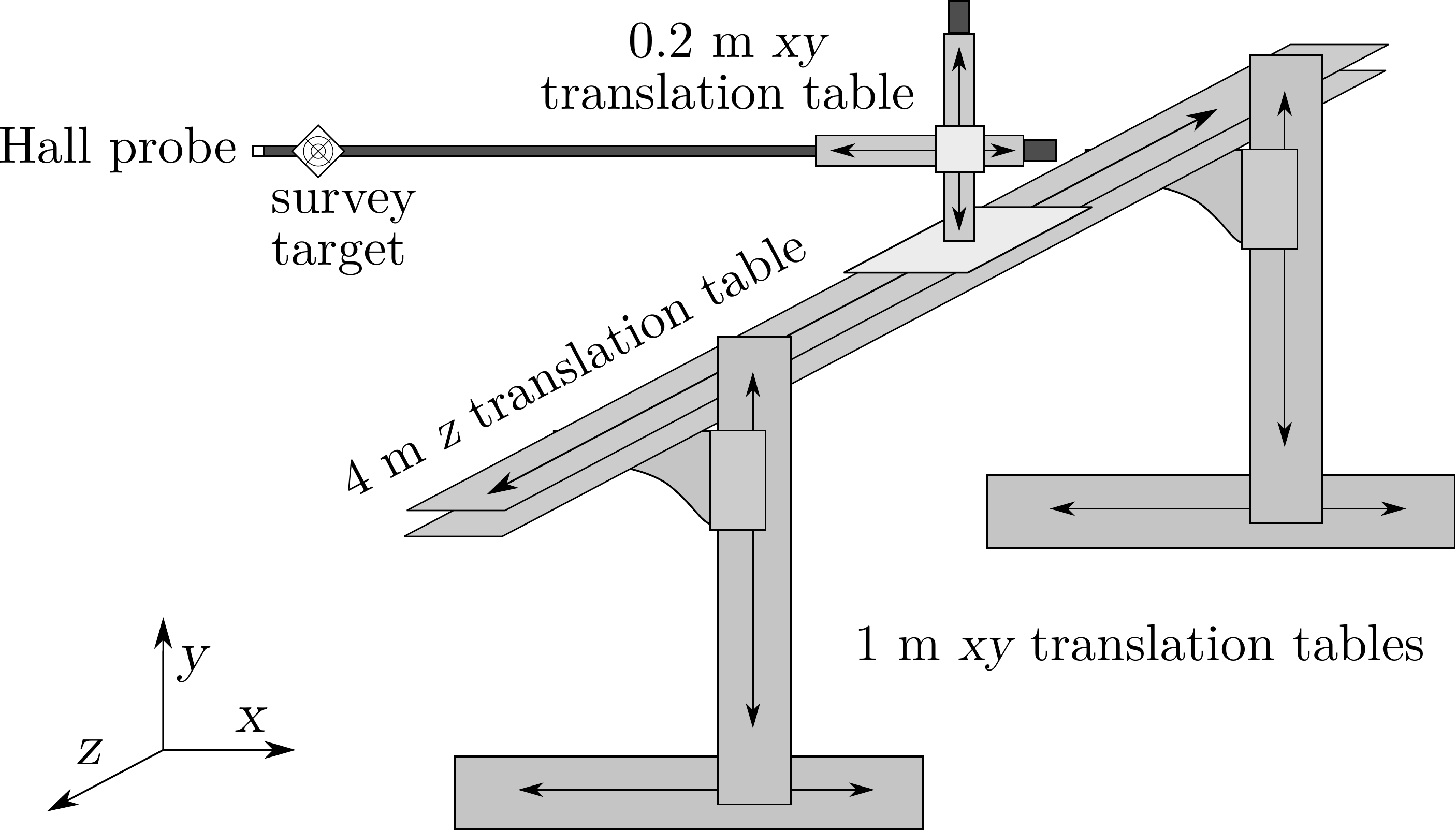}
\caption{Two 1~m $xy$ tables supported a long $xyz$ table, which
could scan 4~m in the $z$ direction. In this configuration, the 
apparatus is assembled to measure the field on the $x>0$ side of 
the toroid.
\label{fig:tables}}
\end{figure}

The measurement apparatus was built from a system of translation tables,
a schematic of which is shown in Figure \ref{fig:tables}. The apparatus
was originally built to measure the field of the undulator magnets of the 
FLASH free electron laser at DESY \cite{Grimm2010105}.
The entire apparatus was supported by a pair of two-dimensional translation
stands, which had 1~m of range in the $x$ and $y$ directions. These stands
were moved synchronously and acted as a single table. This table supported
a three-dimensional translation table with 4~m of range in the $z$ direction
and 0.2~m of range in the $x$ and $y$ directions. This system of translation 
tables was used to move a three-dimensional Hall probe at the end of a carbon fiber rod,
held parallel to the $x$ axis. The range of motion in $x$ and $y$ was 
extended beyond the $1.2$~m range of the translation tables by using rods of different
lengths and different brackets to connect the rods to the tables. 
To allow the Hall probe and rod to move through the magnet volume without
obstructions, the detectors and parts of the beamline were removed before
the apparatus was assembled. 

The position of the Hall probe was monitored using a theodolite equipped 
with a laser range-finder. The theodolite could determine the relative position
to a reflective target, which was attached to the measurement rod. The theodolite's
position was calibrated daily by surveying a set of reference points 
on the walls of the experimental hall and on the frame of the magnet, the 
positions of which were established by previous surveys. 

\begin{figure}[hptb]
\includegraphics[width=\columnwidth]{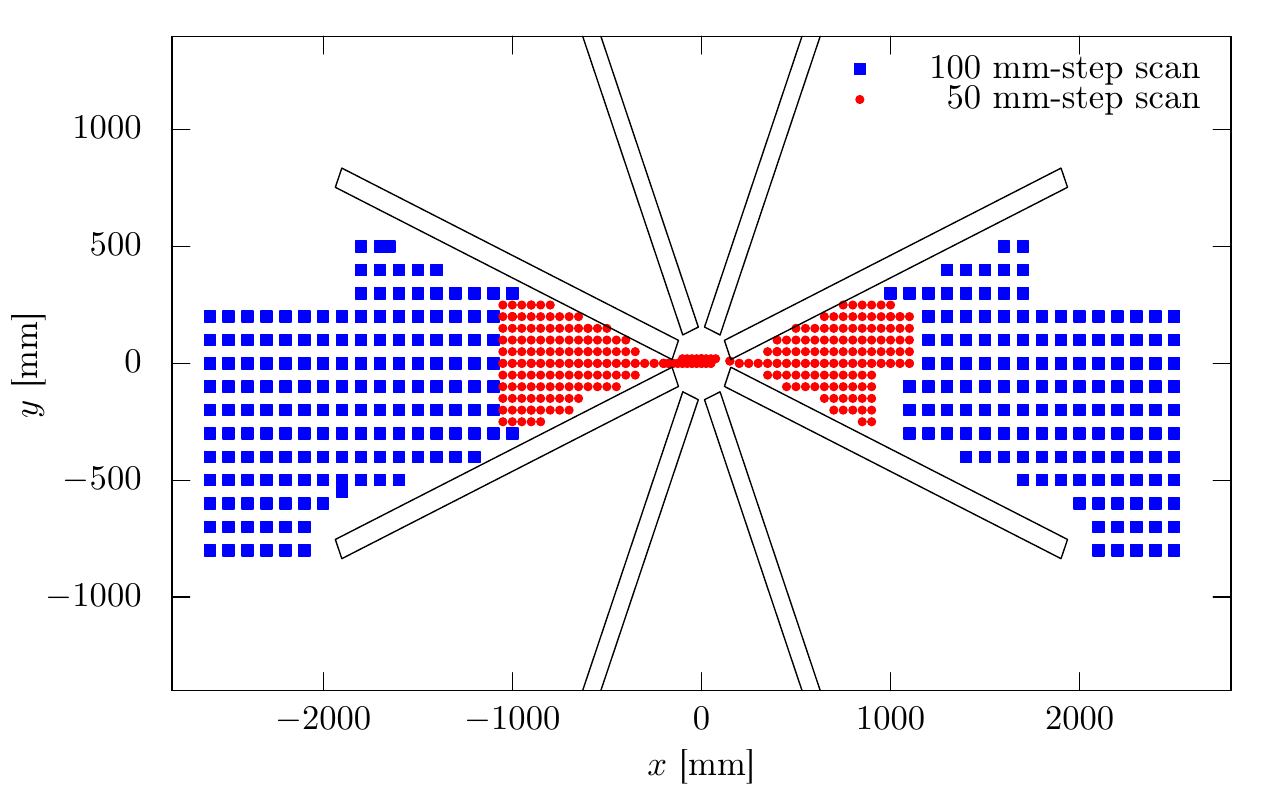}
\caption{Magnetic field measurements were made with greater density in the
inner part of the tracking volume where the field gradients are highest.
\label{fig:scan}}
\end{figure}

Measurement scans were made by moving to a desired $x$ and $y$ position 
and then stepping along the $z$ direction. At the starting and ending point
of a scan, the theodolite was used to survey the position of the 
reflective target. At each step in the scan, the probe would be moved to the 
new $z$ position, followed by a pause of one second to allow any vibrations in 
the rod to dampen. Then, a measurement was made, and the probe was stepped to the 
next point. This procedure was computer-controlled using a LabVIEW application. 
Measurements were made in a three-dimensional grid with 50~mm spacing within 1~m 
of the beamline and 100~mm spacing elsewhere. Measurement scans were made along as 
many $x,y$ trajectories as could be reached, given the ranges of the rods and tables, 
while still avoiding collisions between the magnet and the measurement apparatus. 
The nominal probe positions for each scan are shown in Figure \ref{fig:scan}.
The field was measured at over 36,000 points across both sectors. 

After surveying the left sector, the apparatus was taken apart and reassembled 
for measurements of the right sector. The field in the beamline region was measured from both the 
left and right. These overlapping measurements were consistent in all three field
directions at a level better than $5\times 10^{-5}$~T.

\begin{figure}[hptb]
\includegraphics[width=\columnwidth]{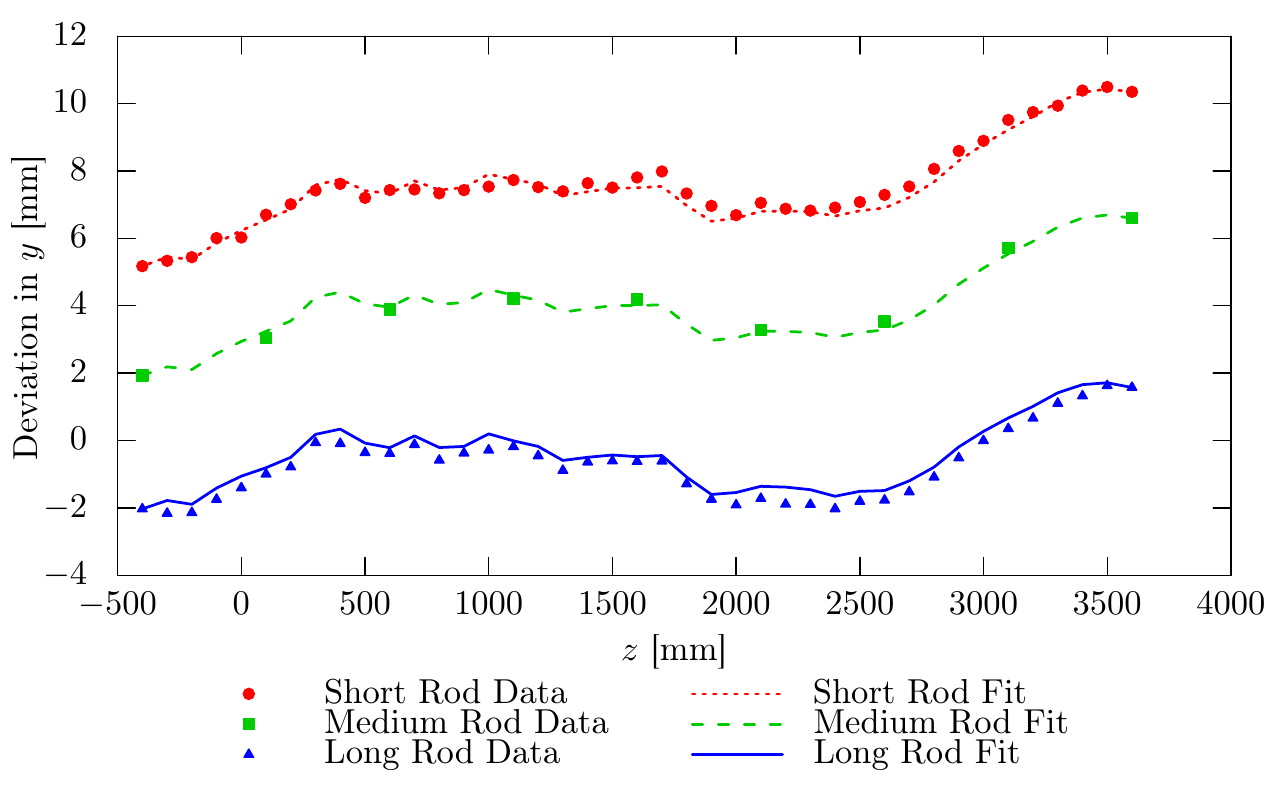}
\caption{The translation tables deviated from their nominal positions over the length of a scan, 
but these deviations could be fit with a simple parameterization. \label{fig:traj}}
\end{figure}

Measurements with the theodolite confirmed that the translation tables did not
provide the millimeter-level accuracy desired. Over the course of $4$~m of translation
in $z$, the long table introduced perturbations of several millimeters in $x$ and $y$.
The position measurements from the theodolite were used to correct for these 
perturbations and recover the true position of the Hall probe. Every time a new rod 
was installed on the tables, a calibration scan was made, in which the reflective 
target was surveyed at every step in the $4$~m $z$ translation. This allowed us to 
determine the three-dimensional trajectory of the target. We found that the trajectories 
had similar shapes for all rods, and the motion could be parameterized with:
\begin{align}
x(z) =& x_t(z) + L_r \cos\left(\theta_t (z)\right) + \text{linear term} \\
y(z) =& y_t(z) + L_r \sin \left(\theta_t (z)\right)  + \text{linear term},
\label{eq:perturb}
\end{align} 
where $x_t$, $y_t$, and $\theta_t$ represent perturbation functions that are
common to the table (independent of which rod was used), and $L_r$ is the length
of the specific rod in use. A linear term was used to match the start and end
points of the trajectory to the starting and ending positions of each measurement scan,
as surveyed by the theodolite. Figure \ref{fig:traj} shows data from three 
calibration scans fit using this parameterization.

\section{Coil Fitting}
\label{sec:coilfitting}

The magnetic field measurement data were fit using a numerical magnetic field 
model, thereby solving two problems. First, the model can be used to calculate
the magnetic field in regions where measurements could not be made, i.e., where the 
probe was obstructed. Secondly, any measurements with errant field readings or offset
position assignments have their local influence dampened in the model. 

In the magnetic field model, the current in each copper winding was approximated 
by a sequence of line segments. The magnetic field produced by a line segment of 
current was found by integrating the Biot-Savart law. The magnetic field at 
position $\vec{x}$ produced by a line segment starting at position $\vec{p}_1$ and 
ending at $\vec{p}_2$, carrying current $I$, is given by:
\begin{equation}
\vec{B} = \frac{\mu_0 I}{4\pi} \frac{\vec{L}\cdot(\hat{V}_2 - \hat{V}_1)}{L^2 
  (\vec{c}-\vec{x})\cdot(\vec{c}-\vec{x}-\vec{L}/L)} \vec{L} \times (\vec{c}-\vec{x})
\end{equation}
\begin{equation}
\vec{c} = (\vec{p}_1 + \vec{p}_2)/2
\end{equation}
\begin{equation}
\vec{L} = \vec{p}_2 - \vec{p}_1
\end{equation}
\begin{equation}
\hat{V}_i = \frac{\vec{p}_i - \vec{x}}{|\vec{p}_i - \vec{x}|}.
\end{equation}

The OLYMPUS coils have straight sections, easily described by line segments, and curved sections,
in which the curves are approximately circular arcs. We approximated the arcs in the model by 
connecting multiple line segments to form a polygon. To divide an arc subtending angle $\alpha$
into $N$ segments, one can place $(N+1)$ vertices evenly along the arc, and connect them to form
a polygon. However, we found we could better match the magnetic field of the arc by placing the 
vertices slightly outside of the arc, so that the polygon and arc had equal area, and thus equal 
dipole moments. In our approximation, we chose to start the first segment at the 
beginning of the arc, and to end the last segment at the end of the arc to maintain continuity. 
We displaced the $(N-1)$ intermediate vertices radially outward from the arc radius $R$ to $R'$,
given by:
\begin{equation}
R' = \frac{R}{N-2}\left[ \sqrt{1 + \frac{\alpha (N-2) }{\sin \left( \frac{\alpha}{N} \right)}} - 1 \right].
\end{equation}

To fit the magnetic field model to the measurements, several parameters were allowed to vary, and a best 
fit was found by minimizing the sum of the squared residuals $\sum |\vec{B}_\text{meas.} - \vec{B}_\text{model}|^2$. 
Several attempts were made in order to strike a balance between giving the model sufficient freedom
to explain the data and introducing degrees of freedom that were unconstrained by the measurements.
Ultimately, a model with 35 free parameters was chosen. The four coils that immediately surrounded 
the measurement region were each given six degrees of freedom (three for positioning and three 
for orientation). The remaining four coils were positioned around a common toroid axis (three parameters
to specify the origin and three parameters to specify the orientation), but were allowed to vary collectively
in radius and in-plane rotation angle. All of the coils were collectively given two degrees of freedom to
stretch or compress in both in-plane directions, since the coils were observed to deform slightly when
current was passed through them. The final free parameter was the current carried in the magnet, which reproduced
the measured input current to within 0.1\%.
In the final fit, arcs were divided into one line segment per degree of curvature. 

\begin{figure}[hptb]
\includegraphics[width=\columnwidth]{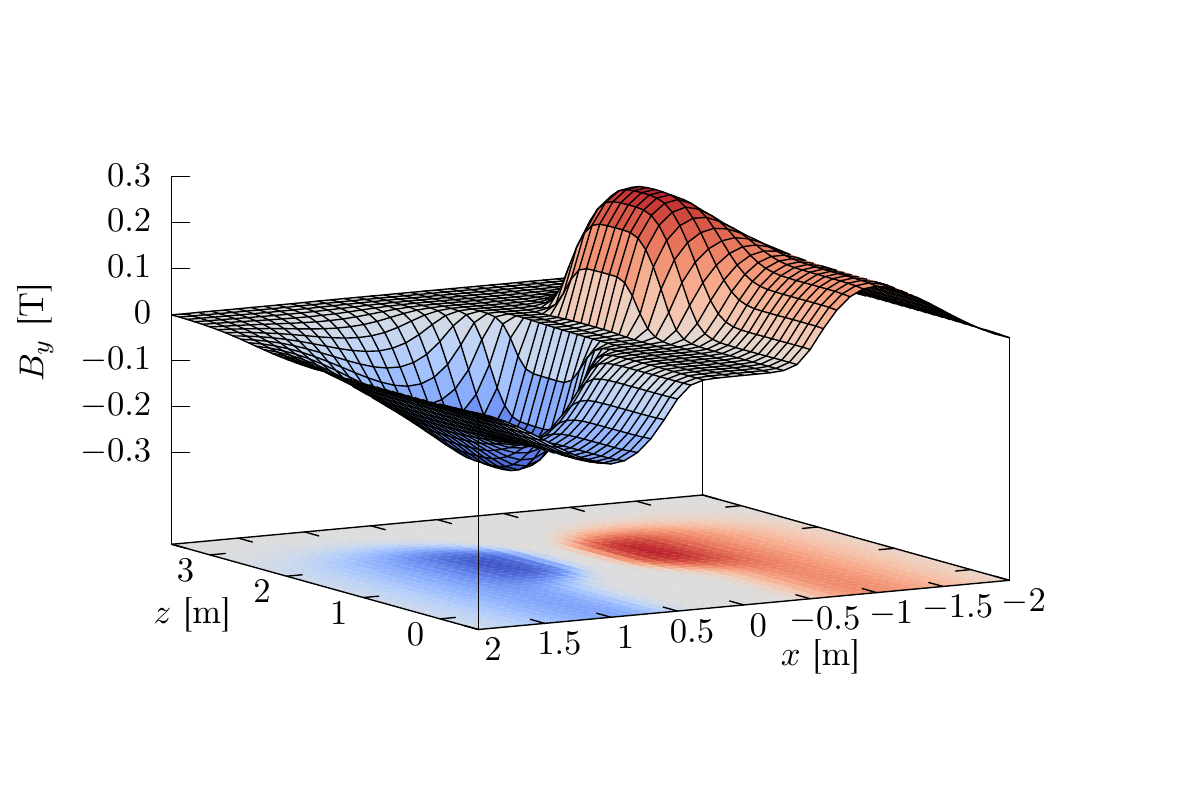}
\caption{The magnetic field in the $y$ direction is shown for the $y=0$ plane. \label{fig:inplane}}
\end{figure}

The model with best-fit parameters had an r.m.s.\ residual: 
$\sqrt{\frac{1}{3N}\sum |\vec{B}_\text{meas.} - \vec{B}_\text{model}|^2}=1.873\times 10^{-3}$~T, where
$N$ describes the total number of measurement points.
The residuals were not uniformly distributed over the measurement region and do not represent 
Gaussian errors. Many systematic effects contribute to the residuals, including any errors in
determining the true probe position and inadequacy of the magnetic field model. The magnetic
field generated by the model with the best fit parameters is shown in Figure \ref{fig:inplane}.

\section{Interpolation}
\label{sec:interpolation}

The model calculation described in the previous section was too slow to be used directly for 
simulating or reconstructing particle trajectories for the OLYMPUS analysis, and so a fast 
interpolation scheme was developed. The coil model was used to pre-calculate the magnetic 
field vector and its spatial derivatives on a regular $50$~mm~$\times$~$50$~mm~$\times$~$50$~mm grid covering the entire 
spectrometer volume, so that the field could be interpolated between surrounding grid points.

The interpolation scheme had to balance several competing goals:
\begin{itemize}
\item Minimizing the memory needed to store the field grid
\item Minimizing computation time for field queries
\item Faithfully reproducing the coil model in both the field and its derivatives.
\end{itemize}
To achieve this, an optimized tricubic spline interpolation scheme was developed based
on the routine of Lekien and Marsden \cite{NME:NME1296}.  For each point $P$ in the grid,
24 coefficients were calculated using the coil model (8 per component of the vector magnetic field):
\begin{multline}
C_{i,P} = \left\{ B_i, \pd{x}{B_i}, \pd{y}{B_i},\pd{x\partial y}{B_i},\pd{z}{B_i},\pd{x\partial z}{B_i},\pd{y\partial z}{B_i},\pd{x\partial y \partial z}{B_i}\right\}
\\ \text{for} \:\: i\in \left\{ x,y,z \right\},
\end{multline}
using numerical differentiation to compute the required derivatives from the coil model.

\begin{figure}[hptb]
\includegraphics[width=\columnwidth]{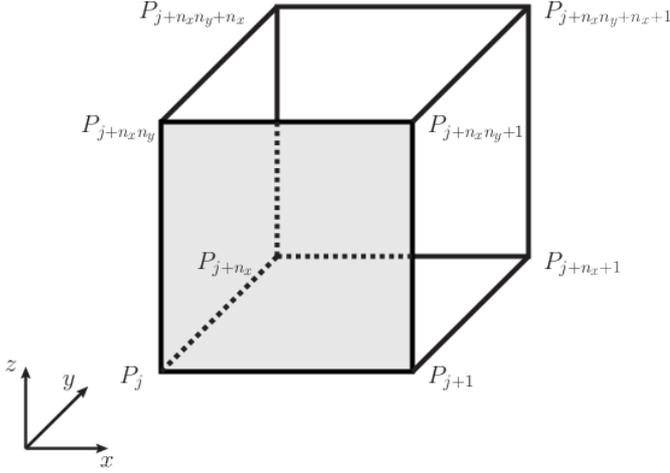}
\caption{A generalized box in the interpolation grid identified by its lowest-indexed grid point $P_j$, where $n_x$ and 
$n_y$ are the $x$ and $y$ dimensions of the grid in units of grid points. \label{fig:grid}}
\end{figure}

For the interpolation, it is convenient to consider the grid in terms of boxes defined
by eight grid points, as shown in Figure \ref{fig:grid}, and define box-fractional 
coordinates $x_f,y_f,z_f \in [0,1] $ parallel to the global axes spanning each box.  
Each point in the grid is labeled with an integer index $j$, such that stepping from
point $P_j$ one unit in $x$ reaches point $P_{j+1}$. Stepping one unit in $y$ from
point $P_j$ reaches $P_{j+n_x}$, where $n_x$ is the size of the grid in the $x$ direction.
Stepping from point $P_j$ one unit in $z$ reaches point $P_{j+n_xn_y}$, where $n_y$ is the 
size of the grid in $y$ direction. Then, a local tricubic spline can be defined for each 
field component in the box:
\begin{equation}
B_i(x,y,z) = \sum_{l,m,n=0}^3 a_{i,lmn} x_f^ly_f^mz_f^n \:\:\:\: i\in \left\{ x,y,z \right\},
\end{equation}
where the coefficients $\left\{a_{i,lmn}\right\}$ are functions of
the set of the 64 parameters $\left\{C_{i,P}\right\}$, where $P$ is any of the eight grid points at 
the vertices of the box.  
This function is a 64-term polynomial for each box and is $C^1$ at the box boundaries. 
The coefficients $\left\{a\right\}$ can be computed from the parameters $C_{i,P}$ following the 
prescription in Lekien and Marsden. However, this prescription requires three $64\times64$ matrix 
multiplications per box. Once completed for a given grid box, these multiplications can be stored 
for future use, but this adds to the size of the grid in memory, approaching a factor of 8 for
large grids.

To avoid these costs, the spline was refactored so that the parameters $C_{i,P}$ can be used 
directly as coefficients. We introduce the following basis functions:
\begin{gather}
f_0(x_i) = \left(x_i-1\right)^2 \left(2x_i+1\right) \\
f_1(x_i) = x_i\left(x_i-1\right)^2 \\
f_2(x_i) = x_i^2\left(3-2x_i\right) \\
f_3(x_i) = x_i\left(x_i-1\right)
\end{gather}
where $ i\in \left\{ x_f,y_f,z_f \right\}$.  The interpolation then takes the form:
\begin{equation}
\label{eq:spline}
B_i(x,y,z) = \sum_{l,m,n=0}^3 b_{i,lmn} f_l(x_f) f_m(y_f) f_n(z_f) \:\:\:\: i\in \left\{ x,y,z \right\},
\end{equation}
where each coefficient $\left\{b_{i,lmn}\right\}$ is one of the parameters $C_{i,P}$. The correspondence
between $\left\{b_{i,lmn}\right\}$ and $C_{i,P}$ is shown in Table \ref{tab:coef}.

\begin{table}[hptb]
\tabcolsep=0.15cm
\begin{center}
\begin{tabular}{l | c c c c c c c c}
   &  $B_i$  &  $ \pd{x}{B_i}$  &  $ \pd{y}{B_i}$  &  $\pd{x\partial y}{B_i}$  &  $\pd{z}{B_i}$  &  $\pd{x\partial z}{B_i}$  &  $\pd{y\partial z}{B_i}$  &  $\pd{x\partial y \partial z}{B_i}$ \\
\hline
$P_{j}$  &  000  &  100  &  010  &  110  &  001  &  101  &  011  & 111 \\
$P_{j+1}$  &  200  &  300  &  210  &  310  &  201  &  301  &  211  & 311 \\
$P_{j+n_x}$  &  020  &  120  &  030  &  130  &  021  &  121  &  031  & 131 \\ 
$P_{j+n_x+1}$  &  220  &  320  &  230  &  330  &  221  &  321  &  231  & 331 \\
$P_{j+n_xn_y}$  &  002  &  102  &  012  &  112  &  003  &  103  &  013  & 113 \\ 
$P_{j+n_xn_y+1}$  &  202  &  302  &  212  &  312  &  203  &  303  &  213  & 313 \\
$P_{j+n_xn_y+n_x}$  &  022  &  122  &  032  &  132  &  023  &  123  &  033  & 133 \\
$P_{j+n_xn_y+n_x+1}$  &  222  &  322  &  232  &  332  &  223  &  323  &  233  & 333 \\
\end{tabular}
\end{center}
\caption{Mapping of the coefficients $\left\{b_{i,lmn}\right\}$ (defined in Equation
\ref{eq:spline}) to the field values and
derivatives at the grid points contained in the box with lowest-indexed point $P_j$.  Entries
in the table are the values of $lmn$ corresponding to each combination of point and
coefficient on the interpolation box. \label{tab:coef}}
\end{table}

With this interpolation scheme, the procedure for querying the field map consisted
of determining the grid box containing the queried point, computing the box fractional
coordinates of the queried point in that box, and then applying the tricubic spline
interpolation for each of the three field components independently.  Special care was
taken to optimize the speed and number of arithmetic operations in the routine (e.g., by pre-computing factors
such as the basis functions that are used repeatedly and by converting division operations to
multiplications whenever possible). Additionally, the coefficients for each grid
point were arranged in a single array so that, for any box in the grid, the 192 coefficients
associated with the 8 points of the box occurred in 16 continuous blocks of 12 coefficients,
permitting rapid reading of the entire array and facilitating single instruction, multiple data (SIMD)
computing, further increasing the speed of field queries.

\section{Special Considerations}
\label{sec:special}

Simulations revealed two regions where the magnetic field model and interpolation were inadequate. 
These special cases are described in the following subsections.

\subsection{Field near the coils}

Some of the interpolation grid points sat very close to line segments of current in the field 
model, which was problematic since the field diverged there. The field and the field derivatives
at these grid points were unreliable, and this spoiled the magnetic field over the eight grid boxes
surrounding every such point. When simulating trajectories that pass close to the coils, we observed that particles had 
abruptly different trajectories if they entered a spoiled grid box. The instrumented region in 
OLYMPUS extends to approximately $\pm 15^\circ$ in azimuth about the $y=0$ plane; aberrant 
trajectories were observed at an azimuth of only $12^\circ$. The problematic points sit near
the coils, at $\pm 22.5^\circ$. 

To safeguard against this problem, we produced a second field grid using a different procedure. 
We defined a safe region, $\pm 15^\circ$ in azimuth from the $y=0$ plane, inside which we trusted
the magnetic field model. (Points inside this region were sufficiently far from the coils to avoid
problems with the diverging field.) For grid points in this region,
the field and its derivatives were calculated as before. For points outside this region, we used
an alternative calculation, exploiting the approximate azimuthal symmetry of the magnet. For each
outside grid point, we first calculated the field and its derivatives at the point with the same 
$z$ and same $r$, but on the nearest $\phi=\pm 15^\circ$ plane. Derivatives were calculated in 
cylindrical coordinates, and any with respect to $\phi$ were set to 0. We then rotated the field 
and derivative vectors back to the point of interest. 

For example, given a grid point at $\phi=20^\circ$, $r=1$~m, $z=2$~m, the field
would first be calculated at $\phi=15^\circ$, $r=1$~m, $z=2$~m. The derivatives $d\vec{B}/dr$,
$d\vec{B}/dz$, and $d^2 \vec{B}/drdz$ would be calculated numerically. All other derivatives
would be set to 0. The vectors would then be rotated by $5^\circ$ in $\phi$, so as to correspond 
appropriately for the grid point at $\phi=20^\circ$, $r=1$~m, $z=2$~m. 

The grid produced using this procedure was interpolated with the scheme in Section \ref{sec:interpolation}. 
Subsequent tests showed that simulated trajectories were not aberrant, even out to $\phi=\pm 15^\circ$,
the limits of the OLYMPUS instrumented region. Furthermore these trajectories were essentially
the same as those simulated with the field model directly, without interpolation.

\subsection{Beamline region}

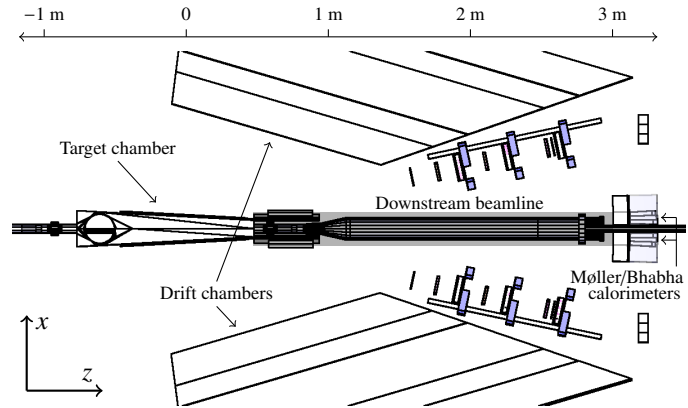
\begin{figure}[hptb]
  
  \begin{tikzpicture}
    \node[anchor=south west,inner sep=0] (image) at (0,0)
         {
           \includegraphics[width=\columnwidth]{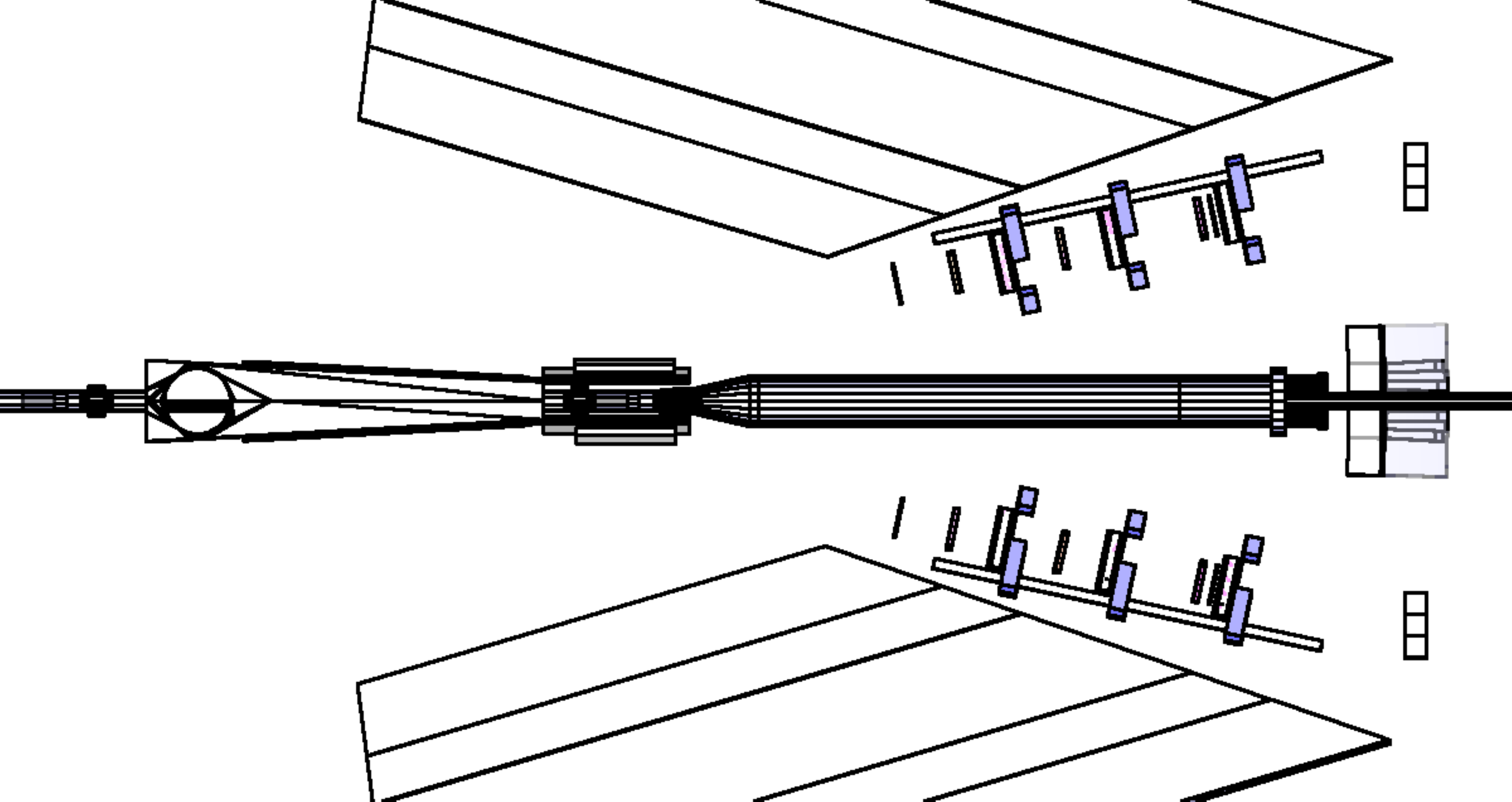}
         };
         \begin{scope}[x={(image.south east)},y={(image.north west)},xscale=.1127,yscale=.212]
           \node[align=center,black] at (0.4,1.1) {$x$};
           \node[align=center,black] at (1.0,0.4) {$z$};
           \node[align=center,black] at (0.43,5.2) {\scriptsize $-1$~m};
           \node[align=center,black] at (2.30,5.2) {\scriptsize $0$};
           \node[align=center,black] at (4.17,5.2) {\scriptsize $1$~m};
           \node[align=center,black] at (6.04,5.2) {\scriptsize $2$~m};
           \node[align=center,black] at (7.91,5.2) {\scriptsize $3$~m};         
           \node[align=center,black] at (2.7,1.5) {\scriptsize Drift chambers};
           \node[align=center,black] at (1.4,3.4) {\scriptsize Target chamber};
           \node[align=center,black] at (5.9,2.7) {\scriptsize Downstream beamline};
           \node[align=center,black] at (8.1,1.7) {\scriptsize M\o ller/Bhabha};
           \node[align=center,black] at (8.2,1.5) {\scriptsize calorimeters};
           \draw[arrows=<->,draw=black] (0.1,4.9) -- (8.5,4.9) node[midway,above]{};
           \draw[arrows=-,draw=black] (0.43,5.0) -- (0.43,4.9) node[midway,above]{};
           \draw[arrows=-,draw=black] (2.3,5.0) -- (2.3,4.9) node[midway,above]{};
           \draw[arrows=-,draw=black] (4.17,5.0) -- (4.17,4.9) node[midway,above]{};
           \draw[arrows=-,draw=black] (6.04,5.0) -- (6.04,4.9) node[midway,above]{};
           \draw[arrows=-,draw=black] (7.91,5.0) -- (7.91,4.9) node[midway,above]{};
           \draw[arrows=->,draw=black] (1.4,3.25) -- (1.9,2.7) node[midway,above]{};
           \draw[arrows=->,draw=black] (2.7,1.65) -- (3.4,3.5) node[midway,above]{};
           \draw[arrows=->,draw=black] (2.7,1.32) -- (2.95,1.05) node[midway,above]{};
           \draw[arrows=-,draw=black] (8.75,1.85) -- (8.75,2.5) node[midway,above]{};
           \draw[arrows=->,draw=black] (8.75,2.2) -- (8.55,2.2) node[midway,above]{};
           \draw[arrows=->,draw=black] (8.75,2.5) -- (8.55,2.5) node[midway,above]{};
           \draw [fill=black, draw=none, opacity=0.3] (3.18,2.13) rectangle (7.91,2.574);
           \draw[arrows=->,draw=black,thick] (0.2,0.2) -- (0.2,1.2) node[midway,above]{};
           \draw[arrows=->,draw=black,thick] (0.2,0.2) -- (1.2,0.2) node[midway,above]{};
         \end{scope}
  \end{tikzpicture}
  \caption{The special beamline field region, shown in the shaded box, contained the entirety of
    the downstream beam pipe from the target cell to the collimators of the M{\o}ller/Bhabha
    calorimeters. \label{fig:symb}}
\end{figure}

The region near the beamline, where the magnetic field was small, was difficult to reproduce 
accurately using the coil fitting model for two reasons. The first is that this region is close
to all eight coils. Slight changes in the model's coil placement can create large gradients 
in the region. The second is that there were few measurements made in that region (and none
inside the volume of the target chamber) to constrain the fit. M{\o}ller and Bhabha tracks 
pass through this region, and an accurate simulation is necessary for the M{\o}ller/Bhabha 
luminosity monitor (described by P{\' e}rez Benito et\ al.~\cite{Benito:2016cmp}). Since the magnetic field model failed in this region,
a dedicated alternative interpolation scheme was developed, for use strictly in this volume.
The region included all points with $\left|x\right|<100$ mm, $\left|y\right|<50$ mm, and $500$~mm~$<z<3000$~mm, shown
in Figure \ref{fig:symb}.

To provide an accurate field map for the forward M{\o}ller/Bhabha scattering region, interpolation was
performed directly on the measured data points in the region. This data was approximately located on the $y=0$ plane
with small variations due to the imperfections of the translation table, described in Section \ref{sec:coilfitting}.  
Since the variation of the field in $y$ was
very small in this region ($\sim$$10^{-4}$~T), the $y$ variation in the grid was ignored and a Lagrange polynomial interpolation 
was used on the remaining irregular 2D grid to produce a regular $5$ mm $\times$ $5$ mm grid in $x$ and $z$ for each of the three
field components in the region \cite{Bevington:1305448}.  This regular grid was then appended to the field map and used for interpolation
of the field in the special region.

During the field measurements, it was not feasible to pass the Hall probe through the walls of the target
chamber. Consequently, no measurements were made for points where $|x| < 100$~mm, $|y|<50$~mm, and $z<500$~mm.
However, a few field measurements were made in 2011, prior to the installation of the chamber, 
and these data confirmed that the field inside the chamber is on the order of $10^{-4}$~T. In this region, 
we chose to use the standard grid interpolation for $B_y$ and $B_z$, since the field model reproduces the 2011 
measurements well in those directions. For $B_x$, a parabolic fit to the 2011 measurements was chosen for
the $x$ dependence, with the assumption that $B_x$ is independent of $y$ and $z$.

\section{Conclusions}

We have described the measurement procedure and the data analysis techniques employed to make a 
comprehensive survey of the OLYMPUS spectrometer's magnetic field. Using an apparatus consisting
of a Hall probe, actuated with a system of translation tables, we measured the magnetic field at
over 36,000 positions in and around the spectrometer volume. We chose to fit these field data 
with a numerical field model to calculate the field at arbitrary positions. 
For analysis applications that required rapid queries of the 
magnetic field, we precomputed the field and its derivatives on a grid of positions, and 
developed a scheme to interpolate the field between grid points using tricubic splines. 
By refactoring the splines, we found that we could reduce the memory needed to store the necessary spline 
coefficients by a factor of eight. This interpolation scheme worked well for the majority of
the spectrometer volume; however, two regions---the region close to the coils, and the region
along the beamline---required special adjustments, which we described. By making these
adjustments, we succeeded in producing a scheme for determining the magnetic field rapidly
and accurately. This is crucial for the OLYMPUS analysis since it allows a high-rate simulation 
of particle trajectories.


\section{Acknowledgements}

We gratefully acknowledge Philipp Altmann at DESY for his assistance in assembling the 
measurement apparatus and preparing for the field measurements. We are also very thankful
for the expertise of Martin Noak at DESY in surveying and aligning the apparatus. 

This work was supported by the Office of Nuclear Physics of the U.S.\
Department of Energy, grant No.\ DE-FG02-94ER40818.





\bibliographystyle{model1-num-names}
\bibliography{references.bib}







\end{document}